# Chemical pressure effect on $T_c$ in BiS$_2$-based Ce$_{1-x}$Nd$_x$O$_{0.5}$F$_{0.5}$BiS$_2$

Joe Kajitani, Atsushi Omachi, Takafumi Hiroi, Osuke Miura, Yoshikazu Mizuguchi

*Tokyo Metropolitan University, 1-1, Minami-Osawa, Hachioji, Tokyo 192-0397, Japan*

Abstract

We have investigated the crystal structure and superconducting properties of the BiS$_2$-based layered superconductor Ce$_{1-x}$Nd$_x$O$_{0.5}$F$_{0.5}$BiS$_2$. Bulk superconductivity was observed for $x \geq 0.4$, and the transition temperature was enhanced with increasing Nd concentration. The highest transition temperature was 4.8 K for $x = 1.0$. With increasing Nd concentration, the length of the *a* axis decreased, while the length of the *c* axis did not show a remarkable change. The chemical pressure along the *a* axis upon Nd substitution seemed to be linked with the inducement of bulk superconductivity. We found that the chemical pressure effect did not completely correspond to the external pressure effect.

Keywords: BiS2-based superconductor; chemical pressure effect

## 1. Introduction

Layered materials have been actively studied in the field of superconductivity because superconductors with unconventional paring mechanisms and/or high transition temperature ($T_c$) had been discovered in layered crystal structures. Recently, we have reported superconductivity in several layered materials possessing a $BiS_2$-type superconducting layer [1,2]. The crustal structure composed of an alternate stacking of the $BiS_2$ superconducting layers and blocking layers is quite similar to those of the Cu-oxide and Fe-based superconductors. So far, three types of $BiS_2$-based materials, $Bi_4O_4S_3$, $REOBiS_2$ family (RE = La, Ce, Pr, Nd, Yb) and $SrFBiS_2$, have found to become superconducting upon electron doping into the Bi-$6p$ orbitals within the $BiS_2$ layers [1-9]. Electrical resistivity measurements under high pressure revealed that the $T_c$ of $BiS_2$-based family was sensitive to application of pressure and can be significantly enhanced [10-12] as observed in the Fe-based family [13].

In the $CeOBiS_2$, bulk superconductivity has not appeared under ambient pressure, but appeared under high pressure with $T_c$ of 8 K. While in the $NdOBiS_2$, the bulk superconductivity has appeared under ambient pressure with $T_c$ of ~5 K. These facts indicate that the superconducting properties of $BiS_2$-based family depend on both the applied pressure and the structure of the blocking layer. In this article, we have investigated the crystal structure and superconducting properties of $Ce_{1-x}Nd_xO_{0.5}F_{0.5}BiS_2$ synthesized to clarify chemical pressure effect on $T_c$ of the $REOBiS_2$ family.

## 2. Experimental

The polycrystalline samples of $Ce_{1-x}Nd_xO_{0.5}F_{0.5}BiS_2$ ($x$ = 0.0 - 1.0) were prepared by a solid-state reaction method. Bi grains, $Ce_2S_3$ powder, $Nd_2S_3$ powder, $BiF_3$ powder, $Bi_2O_3$ powder and $Bi_2S_3$ powder were used as the starting materials. The $Bi_2S_3$ powder was synthesized by a direct reaction of Bi grains and S grains at 500 ºC in an evacuated quartz tube. Other chemicals were purchased from Kojundo-Kagaku laboratory. The mixture of the starting materials with compositions of $Ce_{1-x}Nd_xO_{0.5}F_{0.5}BiS_2$ ($x$ = 0.0 - 1.0) was well-mixed, pelletized and sealed into an evacuated quartz tube. The $Ce_{1-x}Nd_xO_{0.5}F_{0.5}BiS_2$ pellets were heated at 700 ºC for 10h. The obtained products were ground, sealed into an evacuated quartz tube and heated again with the same heating conditions to homogenize the samples. The obtained samples were characterized by X-ray diffraction using the $\theta$–$2\theta$ method. Changes in the lattice volume were discussed with the peak shifts. The temperature dependence of magnetization was measured by a superconducting quantum interface device (SQUID) magnetometer with an applied field of 5 Oe after both zero-field cooling (ZFC) and field cooling (FC).

## 3. Results and discussion

Figure 1 (a) shows the crystal structure of $REOBiS_2$ (RE = Ce, Nd). The X-ray diffraction pattern for the $Ce_{1-x}Nd_xO_{0.5}F_{0.5}BiS_2$ samples is displayed in Fig. 1(b). Almost all of the obtained peaks were explained using the tetragonal $P4/nmm$ space group. The profiles exhibit quite similar tendency upon Nd substitution except for the slight differences in peak shifts corresponding to the lattice contraction. The estimated peak positions of the (200) and (004) peaks are plotted in Figure 2. The shift of the (200) peak to a higher angle corresponds to the shrinkage of the $a$ axis, while that of (004) peak corresponds to the shrinkage of the $c$ axis. Therefore, in this system, the $a$ axis decreases with increasing Nd concentration, while $c$ axis does not show a remarkable dependence on Nd concentration. Figure 3 (a) shows the temperature dependence of magnetic susceptibility for $Ce_{1-x}Nd_xO_{0.5}F_{0.5}BiS_2$. Figure 3(b) shows the Nd concentration dependence of the transition temperature ($T_c$) for $Ce_{1-x}Nd_xO_{0.5}F_{0.5}BiS_2$ estimated from the magnetization measurements. Superconductivity is observed for $x \geq 0.4$. The transition temperature ($T_c$) is defined as a temperature at which the magnetic susceptibility begins to decrease. The $T_c$ increases with increasing Nd concentration and reach 4.8 K for $x$ = 1.0, $NdO_{0.5}F_{0.5}BiS_2$.

The chemical pressure effect induces the lattice shrinkage along the $a$ axis. Bulk superconductivity is observed for $x \geq 0.4$ where the $a$ axis is significantly decreased. The contraction of the $a$ axis seems to be linked with the appearance of superconductivity in the $Ce_{1-x}Nd_xO_{0.5}F_{0.5}BiS_2$ system. However, the obtained $T_c$ is clearly lower than that observed in the external pressure studies on $T_c$ of $CeO_{1-x}F_xBiS_2$. The pressure dependence of $T_c$ in $CeO_{0.5}F_{0.5}BiS_2$ shows a transition-like behavior and exceeds 6 K under high pressure above 2.61 GPa [11]. In fact, the chemical pressure effect on $T_c$ in $CeO_{0.5}F_{0.5}BiS_2$ does not completely correspond to the external pressure effect. Therefore, further increase of $T_c$ at ambient pressure in the $Ce_{1-x}Nd_xO_{0.5}F_{0.5}BiS_2$ system should require optimization of some crystal structure parameters other than the simple contraction of the $a$ axis.

4. Conclusion

In this study, we have systematically synthesized the $Ce_{1-x}Nd_xO_{0.5}F_{0.5}BiS_2$ polycrystalline samples, and investigated the crystal structure and superconducting properties. On the basis of systematic investigation on the lattice contraction and the susceptibility, we found that chemical pressure effect induced lattice shrinkage along the $a$ axis. Bulk superconductivity was observed for $x \geq 0.4$, and the transition temperature was enhanced with increasing Nd concentration. In this system, the chemical pressure along the $a$ axis upon Nd substitution seemed to be linked with the inducement of bulk superconductivity. However, the chemical pressure effect on $T_c$ in $CeO_{0.5}F_{0.5}BiS_2$ does not completely correspond to the external pressure effect. To clarify the detailed correlation between superconductivity and crystal structure, studies using single crystals are required.

Acknowledgement

This work was partly supported by a Grant-in-Aid for Scientific Research for Young Scientists (A) and The Thermal and Electric Energy Technology Foundation.

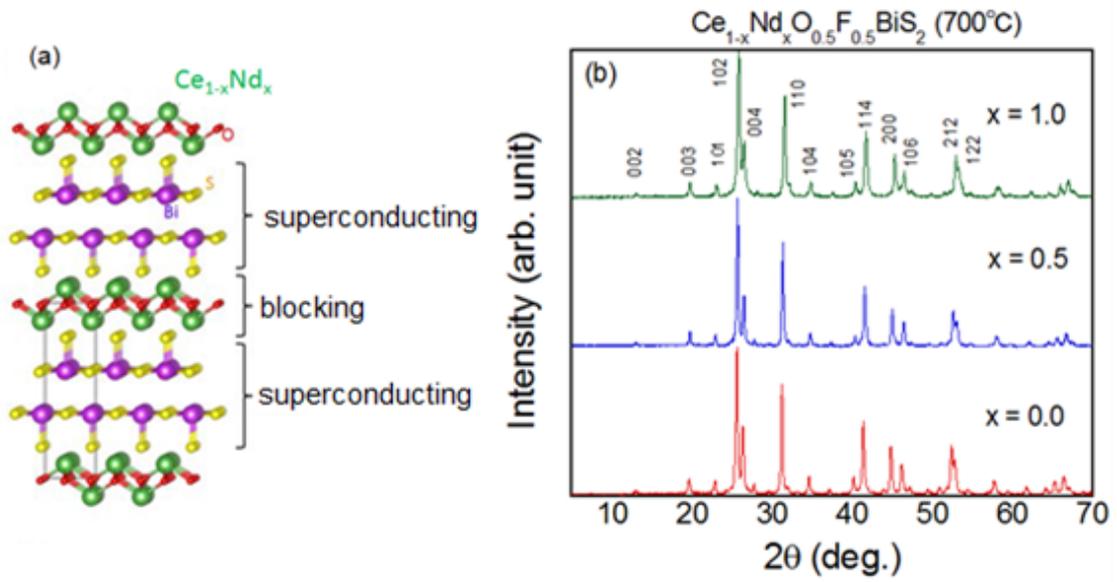

Figure 1. (a) Crystal structure of REOBiS$_2$ (RE = Ce, Nd). (b) X-ray diffraction pattern for Ce$_{1-x}$Nd$_x$O$_{0.5}$F$_{0.5}$BiS$_2$. The numbers indicate Miller indices.

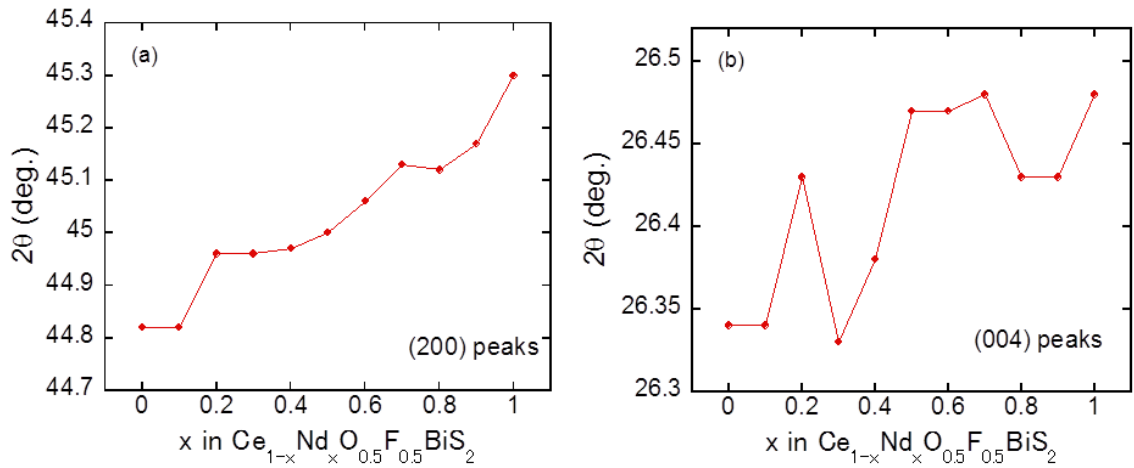

Figure 2. Nd concentration dependence of the peak positions of the (a) (200) and (b) (004) peaks.

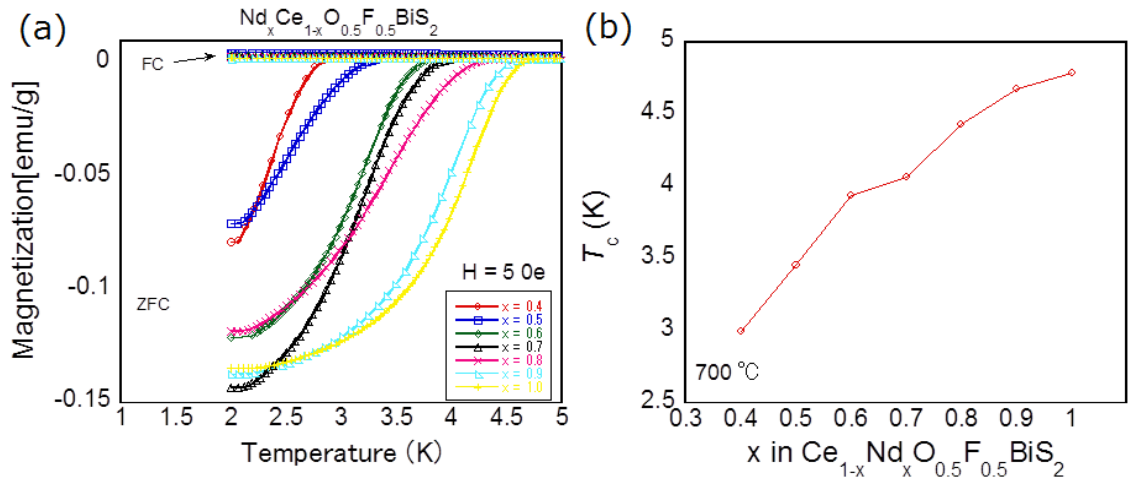

Figure 3. (a) Temperature dependence of magnetic susceptibility for $Ce_{1-x}Nd_xO_{0.5}F_{0.5}BiS_2$. (b) Nd concentration dependence of the transition temperature ($T_c$) for $Ce_{1-x}Nd_xO_{0.5}F_{0.5}BiS_2$.